\documentclass[a4paper]{article}
\usepackage{amsmath,amstext,amsgen,amsbsy,amsopn,amsfonts,amsthm,amssymb}
\usepackage{easybmat}
\usepackage{graphics}
\usepackage[pdftex]{graphicx}
\usepackage{wrapfig,epsfig}
\usepackage{bm}
\usepackage{algorithm}
\usepackage{algorithmic}
\usepackage{esvect}
\usepackage{multirow}
\usepackage{array}
\usepackage{booktabs}
\usepackage{xcolor}
\usepackage{hyperref}
\usepackage{fancyvrb}
\usepackage{gensymb}

\providecommand{\keywords}[1]{\textbf{\textit{Index terms---}} #1}

\begin{document}
\title{\textbf{Cost-sensitive Selection of Variables by \\Ensemble of Model Sequences}}
\author{
Donghui Yan$^{\dag\P}$, Zhiwei Qin$^{\$}$, Songxiang Gu$^{\S}$, \\Haiping Xu$^{\ddag\P}$, Ming Shao$^{\ddag\P}$
\vspace{0.12in}\\
$^\dag$Department of Mathematics and Program in Data Science\vspace{0.05in}\\
$^\$$DiDi Research America, Mountain View, CA\vspace{0.05in}\\
$^\S$JD Digital, Mountain View, CA\vspace{0.05in}\\
$^\ddag$Department of Computer and Information Science\vspace{0.05in}\\
$\P$University of Massachusetts Dartmouth, MA\vspace{0.05in}
}

\date{\today}
\maketitle

\begin{abstract}
\noindent
Many applications require the collection of data on different variables or measurements over many system performance 
metrics. We term those broadly as measures or variables. Often data collection along each measure incurs a cost, thus it is desirable
to consider the cost of measures in modeling. This is a fairly new class of problems in the area of cost-sensitive learning. A few attempts 
have been made to incorporate costs in combining and selecting measures. However, existing studies either do not strictly enforce 
a budget constraint, or are not the `most' cost effective. With a focus on classification problems, we propose a computationally 
efficient approach that could find a near optimal model under a given budget by exploring the most `promising' part of the 
solution space. Instead of outputting a single model, we produce a {\it model schedule}---a list of models, sorted by model costs 
and expected predictive accuracy. This could be used to choose the model with the best predictive accuracy under a given budget, 
or to trade off between the budget and the predictive accuracy. Experiments on some benchmark datasets show that our 
approach compares favorably to competing methods.
\end{abstract}


\section{Introduction}
\label{section:Intro}
Many applications require the collection of data on different variables or measurements over a number of
system performance metrics. For example, some cyber systems rely on scanning various system 
metrics to detect or to predict potential cyber intrusions or threats. In the maintenance of airplanes or major factory 
machinery, measurements of different system components and their usage statistics are collected to determine when 
a maintenance is required. In medical diagnosis, a patient may be asked to take various medical tests on measures 
such as blood pressure, cholesterol level, heart rates and so on, so that the doctor could determine if 
the patient has a certain disease. In the development of an e-commerce product that predicts 
the click or purchase of a product at an e-commerce website, many data related to a 
user's shopping behavior will be collected, and often extra data relevant to the product or the user's shopping behavior 
are purchased from a third-party vendor etc. The data collected on various measures need to be combined, and if 
cost is a concern, a subset of measures needs to be selected to satisfy the budget constraint.  
\\
\\
The problem of combining measures for a target application can be formulated as follows. Assume there are $p$ measures,
then a measurement of the system will be a vector in $\mathbb{R}^p$. Let $X_i=(X_{i1},X_{i2}, ...,X_{ip})$ be an instance
of {\it measurement} with $X_{ij}$ indicating the $i^{th}$ measurement on the $j^{th}$ measure. Each measurement is 
associated with a state variable, denoted by $Y_i$, indicating the system status. Examples of the state variable include an 
indicator of whether a person is healthy or otherwise in a health screening or diagnosis, whether a major repair is required 
in airplane or machinery maintenance, whether a cyber system is under attack, or an indicator on the click or purchase of a 
product item in an e-commerce application. By collecting a sample of measurements and the associated status, we can estimate 
their relationship
\begin{equation*}
f: X \mapsto Y.
\end{equation*} 
That is, a model of the system operation---the relationship between the measurement and the system status. Or, for measurement 
$X=x$, what would be the likelihood of a certain event, such as a disease, a cyber attack, or an immediate repair of some 
airplane parts or machinery. Our interest is to solve the prediction or classification problem. Formally, we seek to solve the following 
\begin{equation}
\label{eq: riskMinimization}
\arg\min_{f \in \mathcal{F}} \mathbb{E} l(f(X),Y),
\end{equation}
where $\mathcal{F}$ is the function class of interest, such as linear classifiers, decision trees etc, $l(.,.)$ is 
the loss function, and $\mathbb{E}$ indicates that we are taking expectation over the distribution of $(X,Y)$ 
(i.e., expected risk in future prediction). We consider the 
simplest loss function, the 0-1 loss, for which \eqref{eq: riskMinimization} amounts to solve 
for $f$ for the best predictive accuracy. 
\\
\\
In practice, the measurements along each variable may incur a cost; sometimes the cost may be substantial. Let 
$\bm{b}=(b_1,...,b_p)$ denote the {\it cost profile} where $b_i$ stands for the cost of the $i^{th}$ variable. It is highly 
desirable, sometimes mandated, that the model should be built under a total budget, say, $B$. That is, the total cost for 
variables used by the model satisfies the following constraint
\begin{eqnarray}
&& \sum_{\beta_i \neq 0} b_i \leq B,
\label{eq:costConstraint}
\end{eqnarray} 
where $\beta_i$ is either the coefficient of the $i^{th}$ variable in a linear model, or otherwise an indicator of whether 
the $i^{th}$ variable is present in the model, $i=1,...,p$. We call \eqref{eq: riskMinimization}, with the additional constraint 
\eqref{eq:costConstraint}, the problem of {\it cost-sensitive selection of measures} or {\it variable selection under a budget}, 
and this is the focus of the present work. When the costs of all variables are equal, i.e., $b_1=b_2=...=b_p$, it 
reduces to the usual feature subset selection problem. 
\\
\\
Finding a subset of variables so that they collectively achieve a good predictive accuracy is a 
challenging problem. The major difficulty lies in the fact that it is a discrete optimization problem (or more particularly, 
the {\it cardinality problem})---as for those variables taking discrete values, one cannot apply gradient descent types of 
algorithm and {\it all possible combinations of possible discrete values} will have to be evaluated in order to find the optimal solution. 
When the number of variables increases, it quickly leads to a combinatorial explosion. Clearly the problem becomes 
harder when incorporating a budget constraint on the total cost of selected variables. As a result, 
often solutions resort to heuristics. 
\\
\\
The overall strategy in our proposed approach is to explore the solution space in an efficient 
way. We achieve this by aiming at those {\it critical points} in the solution space, in the sense that such points are 
either themselves `special'  (one could think of such points as vertices of the polytope of feasible solutions in linear 
programming \cite{Luenberger2003}) or would allow us to gauge the values of many others approximately. Thus, if 
one is able to visit those critical points, or come close to such points, then effectively one has explored a large portion 
of the solution space. We implement this by following a number of `promising' solution sequences. Each solution 
sequence consists of a series of solutions to the target optimization problem (i.e., a model to the cost-sensitive 
learning problem) that are organized by simulating the path of a gradient optimization along a certain direction. This 
leads to either a sequence that progressively removes the least predictive variables when starting with the full model 
(that is, a model that uses all the variables), or that sequentially removes the most costly variables, or that iteratively 
removes the least important variables by randomly sampling with a weight inversely proportional to some normalized
importance measure. As the total cost of variables can be calculated for each model, by examining individual 
models in all the model sequences, we can find all solutions among the model sequences that satisfies a predefined 
budget constraint. Then we can learn the predictive accuracy of all such solutions using some validation data, and pick
the one with the best predictive accuracy as the solution to the budget constrained learning problem.  
\\
\\
Our contributions are as follows. We rigorously formulate the problem of learning under a budget constraint on variables. 
Previous work either does not strictly enforce the budget constraint (e.g., \cite{ZhouZhouLi2016}), or is limited to a 
particular learning method (e.g., \cite{L1Budget2018} specializes to $L_1$-logistic regression). Our approach 
would work for any classification method, especially those strong classifiers, such as Random 
Forests (RF) \cite{RF}, boosting and its variants \cite{FreundSchapire1996,Friedman2002,ChenGuestrin2016}.
Our approach also makes it possible to use different classifiers to generate multiple model sequences. As an ensemble 
based method, one would likely see a boost in performance; it is possible to generate individual model sequences 
in the ensemble in parallel by multicore computing \cite{dc22019}. We discover a new use of the $L_1$-regularization 
path as an efficient way of exploring the model space, rather than a necessary part of the model fitting 
procedure; thus one can use $L_1$-logistic regression to generate the regularization path and then apply a different
classifier on the set of variables selected at each step of the regularization path. We generalize the concept of a {\it 
single predefined budget} to that of a model schedule, which specifies what would be the best predictive model under 
each of a list of varying budget levels. This leads to a much greater flexibility in practice. Given a budget, one can 
choose a model from the model schedule that would deliver the best predictive accuracy, or achieve a given prediction 
accuracy level with the least budget, or trade-off between the budget and the expected predictive accuracy. Finally, our idea 
of multi-path search, along with the learning of model power from data and multicore computing, can become a 
general strategy for efficiently finding an approximate solution to a large class of discrete optimization problems. 
\\
\\
The remaining of this paper is organized as follows. In Section~\ref{section:backgroundRelated}, we introduce 
necessary background and discuss related work. This is followed by a detailed description of our proposed approach
in Section~\ref{section:method}. In Section~\ref{section:optimality}, 
we define the optimal model schedule, and use a small scale problem to demonstrate the optimality of the model schedule 
produced by our algorithm. In Section~\ref{section:experiments}, we present experimental results on some real datasets. 
Finally, we conclude in Section~\ref{section:conclusions}. 
\section{Background and related work}
\label{section:backgroundRelated}
In this section, we discuss necessary background and work related to ours. We start by an introduction to two 
machine learning algorithms, RF and $L_1$-logistic regression, which are important ingredients of our algorithm. 
Then we discuss work related to ours. 
\subsection{Random Forests}
\label{section:methodRF}
RF is widely viewed as one of the most powerful tools in statistics and 
machine learning according to many empirical studies \cite{RF,Caruana2006,caruanaKY2008}. 
It is an ensemble of decision trees with each tree constructed on a bootstrap resample of the data. 
Each tree is built by recursively partitioning the data space. At each node (the root node corresponds 
to the bootstrap sample), RF randomly samples a number of features (or sets of features) and then 
select one that would lead to an `optimal' partition of that node. This process continues recursively 
until a stopping criterion is met. \cite{RF} argues RF would achieve an `optimal' bias and 
variance combination by fully growing individual trees (for classification). RF is easy to use (e.g., very 
few tuning parameters) and shows a remarkable built-in ability for feature selection. 
\\
\\
The computational complexity of RF can be calculated as follows. The average height of a tree is given 
by $O(\log(n))$, and each level in the tree involves the calculation of the merit of $O(p^{1/2})$ candidate 
variables on all the observations and the subsequent search of best split, which costs $O(p^{1/2} \cdot n\log(n))$
in total. Assume the total number of trees in RF is $T$, then the average computational complexity in 
growing RF is given by $O(T \cdot p^{1/2}\cdot n(\log(n))^2)$.
Similarly the applying RF on the test or validation set, with a sample size of $n$, costs $O(T\cdot n\log(n))$. 
\subsection{The $L_1$-logistic regression}
\label{section:l1Regression}
\begin{figure}
\centering
\begin{center}
\hspace{0cm}
\includegraphics[scale=0.72,clip,angle=-90]{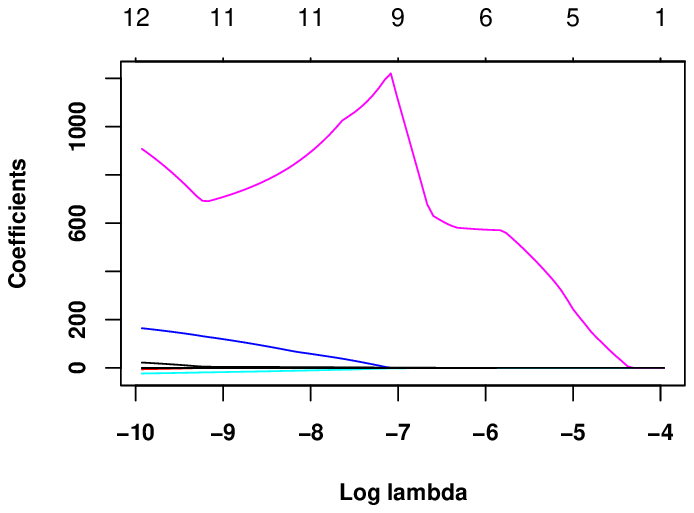}
\end{center}
\abovecaptionskip=-5pt
\caption{\it The regularization path of $L_1$ logistic regression. Each curve corresponds to the coefficient of one variable. 
As $\lambda$ increases, the value of some coefficients will shrink to 0.} 
\label{figure:l1path}
\end{figure}

The $L_1$-logistic regression is the usual logistic regression with an $L_1$-penalty
on the coefficients in the logistic regression model. That is, to solve optimization problem \eqref{eq:logitL1}. 
\begin{equation}
\arg \min_{\beta=(\beta_1,...\beta_p)} L(\beta)=-\log\left(\Pi_{i=1}^n \left[ p(X_i) \right]^{Y_i} \left[ 1-p(X_i) \right]^{1-Y_i} \right)
+ \lambda \sum_{i=1}^p |\beta_i|, 
\label{eq:logitL1}
\end{equation} 
where $\lambda$ is a regularization parameter, $p(X_i) \triangleq P(Y_i=1|X_i)$ is the posterior probability, 
$\beta=(\beta_1,...,\beta_p)$ is a vector of coefficients in the linear model (logistic regression models the 
$\log$-odds ratio of the posterior probability as a linear model of independent variables) and 
$i=1,...,n$, and $n$ is the training sample size. The optimization problem \eqref{eq:logitL1} is typically solved 
by a gradient descent type of algorithm. A variant, coordinate descent, is used by the popular package {\it glmnet()} \cite{glmnet2010},
which does gradient descent along one variable at a time while keeping other variables fixed, and the variables
are evaluated in turn until convergence (after a predefined number, $E$, steps). 
A full range of different values of $\lambda$ are attempted, and each leads to a feasible solution. One then uses 
some model selection procedure, such as cross-validation, to select the $\lambda$ value that would lead to the 
best predictive performance.
\\
\\
{\it The $L_1$-regularization path} is a sequence of solutions to \eqref{eq:logitL1} under different values of $\lambda$ 
such that $\lambda_1 > \lambda_2 > ...>\lambda_{n_{\lambda}}$, where $n_{\lambda}$ depends on the number of steps 
one wishes to include in the regularization path. Typically $\lambda_1$ is chosen such that the model consists of only 
the intercept, $\lambda_{n_{\lambda}}=0$ implies no regularization, and $\lambda_i, i=2,...,n_{\lambda}-1$, are chosen 
adaptively such that their choices will cause a change to current set of variables in the model. For details about model 
fitting in $L_1$-logistic regression, please refer to \cite{ParkHastie2007,glmnet2010}. Each solution to 
\eqref{eq:logitL1} corresponds to a model. A nice property of $L_1$-regularization is the sparsity of the solution, i.e., if 
one keeps on increasing $\lambda$, then the coefficient of some parameters will shrink towards 0. This can be seen in 
Figure~\ref{figure:l1path}.
Thus a regularization path corresponds to an organized sequence of fitted models. 
\\
\\
The computational complexity of $L_1$-logistic regression is calculated as follows. The computing of gradient 
optimization in logistic regression costs $O(p\cdot J\cdot E\cdot n)$, where $E$ is number of epochs in gradient 
descents and $J$ is the number of classes. The full regularization path for $L_1$-logistic regression 
involves $n_{\lambda}$ steps of logistic regression, thus has a computational complexity 
$O(n_{\lambda} \cdot p\cdot J\cdot E\cdot n)$ \cite{ParkHastie2007,glmnet2010}. 
\subsection{Related work}
\label{section:related}
Work related to ours fall into two categories. One is on variable selection, also known as feature selection or model selection. 
The other is on work that incorporates a cost in the model, known generally as cost-sensitive learning. The literature on feature 
selection is enormous, we shall refer the readers to \cite{Akaike1974,Schwarz1978,LiuMotoda1998,GuyonElisseeff2003,PengLongDing2005} 
and references therein for early work. More recent developments include numerous methods based on the idea of regularization \cite{Tibshirani1996,EfronHJT2004,ZouHastie2005,MeinshausenBuhlmann2006,ParkHastie2007,glmnet2010}, feature screening 
\cite{Wang2009,WangLeng2016}, univariate statistics \cite{DonohoJin2008,DelaigleHallJin2011,TangAlelyaniLiu2014} etc. The 
development on feature selection has been explosive during the last decades, and references listed here are just a small sample 
of the huge body of literature.  
\\
\\
The seminal work by \cite{Elkan2001} is an early work on cost-sensitive learning. It is an example that associates
costs with data instances for a classification error. In classification, usually the 
same loss is incurred to each data instance for an error, but \cite{Elkan2001} distinguishes errors committed 
to different classes and charges at different costs. For example, there would be a different cost for errors in classifying 
a safe system to be under attack and those errors in mis-detecting a cyber attack. There are a number of followups 
\cite{ShengLing2006,OBrienGuptaGray2008} and extension of the cost to per example based \cite{ZadronyLangfordAbe2003}. 
Cost-sensitive learning has also been studied in the setting of active classifier \cite{GreinerGR2002} and adaptive 
feature acquisition \cite{JiCarin2007}. 
\\
\\
Incorporating a cost for features is a fairly new area. \cite{ZhouZhouLi2016} considers the cost of variables in RF, 
where, at each node split in the construction of decision trees, variables are selected by sampling with a probability inversely 
proportional to their costs. While the resulting classifier may have a low-cost in variables, it does not necessarily satisfy 
the budget constraint. In cases when those more predictive variables also have a high variable cost, this will badly
hurt the performance of the resulting classifier as one looks for {\it collectively not individually} inexpensive variables. 
The work that is most closely related to ours is \cite{L1Budget2018} which achieves a cost-sensitive combination 
and selection of measures by an {\it $L_1$-logistic regression} formulation. In particular, it incorporates an 
$L_1$-penalty \cite{Tibshirani1996,ParkHastie2007,glmnet2010} in the model fitting of logistic regression 
with an additional cost constraint as follows
\begin{eqnarray}
\label{eq:logitL2}
&& \arg \min_{\beta=(\beta_1,...\beta_p)} L(\beta)=-\log\left(\Pi_{i=1}^n \left[ p(X_i) \right]^{Y_i} \left[ 1-p(X_i) \right]^{1-Y_i} \right)
+ \lambda \sum_{i=1}^p |\beta_i|, \\
&& \sum_{\beta_i \neq 0} b_i \leq B.
\label{eq:budget}
\end{eqnarray} 
Then it navigates through the $L_1$-regularization path for \eqref{eq:logitL2}, and generates a sequence of models 
with each using the $\beta$ corresponding to a different value of $\lambda$ (there are totally $n_{\lambda}$ such $\lambda$'s). 
As $L_1$-regularization encourages sparse models, the solution given by \cite{L1Budget2018} typically yields a 
model with a satisfactory predictive performance at a low cost. The budget constraint \eqref{eq:budget} is enforced by 
following the $L_1$-regularization path, and then the best predictive model (based on performance on a validation set) 
that is under the budget is selected as the solution to the variable selection problem under a budget. While this work
selects variables that enforces the budget constraint, it is limited to $L_1$-logistic regression.
\\
\\
We take a broad view of the $L_1$-regularization path explored in \cite{L1Budget2018}, and view it as an
effective search path in the solution space to the challenging optimization problem for the cost-sensitive variable 
selection problem. This inspires us to consider more search paths other than the $L_1$-regularization path, which 
may potentially overcome its limitations, thus more model sequences can be packed to form an ensemble; this would
lead to a substantial gain in the resulting model. We also extend the scope of budget, thus more flexibility, by proposing
a concept called the {\it model schedule} which is a table of budgets and the associated best predictive model under 
each budget level. Thus, one just needs to run our algorithm once and would then be able to give the best predictive 
model for any particular budget level, or to tell what would be the minimal budget required to deliver a model with a 
predefined predictive accuracy. 
\section{Algorithm for model sequences search under a budget}
\label{section:method}
Our main idea in solving the challenging optimization problem for the selection of variables under budgets 
is to efficiently explore the most promising part of the solution space. This is implemented by visiting multiple 
model sequences, with each having the potential of finding a `near' optimal solution (each model corresponds 
to a point or a feasible solution to the target optimization problem). Such model sequences are typically greedy in 
nature, and are pursued separately in the practice of variable selection which often lead to decent results. Thus
we expect each individual model sequence explored in our search are already `close' to the optimal solution. 
Now combining such model sequences would likely lead to improvement. 
As the resulting model sequences are often a nested sequence of models (i.e., the set of variables in a 
model is a subset of that of preceding models in the sequence), they can all be constructed efficiently. 
We term our approach {\it msB}, short for model sequences search under a budget.
\\
\\
Note that, for each model in a model 
sequence, we can form a tuple by the parameters of this model, along with the model cost and an expected predictive 
accuracy produced from a validation set. We can treat tuples resulting from the same model sequence as a set. This 
would allow one to easily combine model sequences as the union of sets. To make the model sequence directly useable 
in practice, 
we apply two operations. First the model sequence is {\it sorted} by the model cost and predictive accuracy. Then, the 
model sequence is {\it compressed} such that those members with a higher model cost but lower predictive accuracy 
than others will be removed from the model sequence. This produces a {\it model schedule} which is a list of models 
with increasing cost and predictive accuracy. Note that a model schedule has the monotonicity property---higher cost 
models in the schedule always lead to a higher expected predictive accuracy.
\\
\\
For the rest of this section, we will describe different ways in generating model sequences, and how to produce the 
output model schedule from such model sequences. The generation of such model sequences is illustrated by the 
Naval propulsion plants (NPP) dataset (for a brief description, please see Section~\ref{section:experiments}).
\subsection{Generating multiple model sequences}
\label{section:methodMultiSeq}
We consider four different ways of generating model sequences, including following the $L_1$-regularization 
path, selecting variables by their importance, selecting variables by their costs, and sampling according to a 
tradeoff between cost and variable importance. Other ways of generating model sequences, such as forward 
or backward stagewise variable selection, can also be considered. We will discuss each of the four different model sequence 
generation procedures in the sequel.
\subsubsection{Model sequence by importance or cost of variables}
\label{section:methodICS}
\begin{figure}
\centering
\begin{center}
\hspace{0cm}
\includegraphics[scale=0.5,clip,angle=-90]{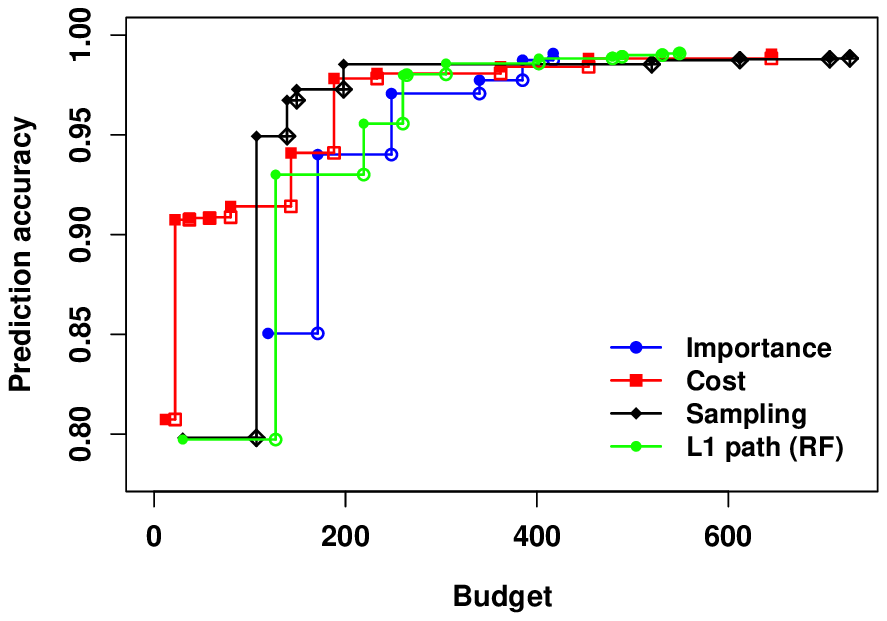}
\end{center}
\abovecaptionskip=-5pt
\caption{\it Model schedules generated by different model sequences, including that by variable importance, by variable
cost, sampling by normalized variable importance, and by following the $L_1$ regularization path. Note that here RF is 
used as the classifier to generate all the model schedules, and we use $``L_1~path~(RF)"$ to emphasize that the classifier is
RF not $L_1$-logistic regression.} 
\label{figure:NPindividuals}
\end{figure}

To produce a nested sequence of models by variable importance, we first rank the variables by their importance
to predictive accuracy. There are many ways around in doing this, for example, by t-statistics \cite{LiuMotoda1998}.
As we use RF as the engine for selecting and combining variables, we will use a built-in tool by RF to produce 
a variable importance profile. 
\begin{figure}
\centering
\begin{center}
\hspace{0cm}
\includegraphics[scale=0.53,clip,angle=-90]{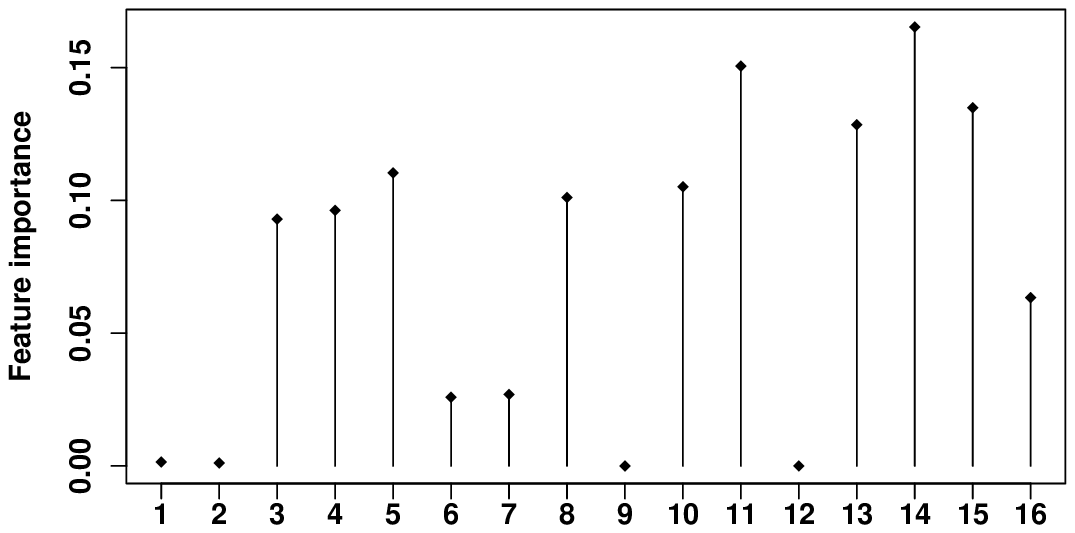}
\end{center}
\abovecaptionskip=-10pt
\caption{\it Feature importance produced by RF on the Naval propulsion plants dataset. The x-axis indicates feature index.} 
\label{figure:costvarImpNP}
\end{figure}
There are two feature importance metrics in RF, 
one based on the Gini index \cite{CART,RF} and the other on permutation accuracy. We consider the latter here, as it 
is often considered superior. The idea is as follows. Randomly permute the values of a feature, say, the $i\textsuperscript{th}$ 
feature, then its association with the response $Y$ is broken. When this feature, along with those un-permuted features, is 
used for prediction, the accuracy tends to decrease. The difference in the prediction accuracy before and after permuting 
the $i\textsuperscript{th}$ feature can then be used as a measure of its importance. Figure~\ref{figure:costvarImpNP} shows 
the relative importance of different variables used in the NPP data.
\\
\\
With a profile of variables importance, a nested sequence models is produced as follows. We start with the {\it full model}, 
that is, a model with all variables present. Then, we delete the least important variable, according to its importance value;
this gives a new model. We record its predictive accuracy on a validation set and compute the total 
cost of all variables in the new model. This procedure continues until there are two variables left (at which point we have 
to stop as RF does not allow less than two variables). This produces a list of models, with such information as model cost, 
predictive accuracy, and variables used. From this list, we can generate a model schedule specifying at which cost, what 
kind of predictive accuracy we can expect. It may happen that there are models in the list with a higher cost but lower predictive 
accuracy---when this happens we simply remove such models from the list. Thus in the final model schedule, 
a higher-cost model would have a higher expected predictive accuracy. The following is an instance of model schedule 
under a certain cost profile (i.e., $\bm{B}=(b_1, b_2, ..., b_p)$ where $b_i$ is the cost of the $i\textsuperscript{th}$ variable) 
and the variable importance profile shown in Figure~\ref{figure:costvarImpNP}. Here the cost profile is 
generated by sampling uniformly at random from $[1,100]$; the same applies to all figures in this section.
\\
\begin{Verbatim}[fontsize=\small]
       Cost    Accuracy       Variables
[1]    417     0.9907834     4,5,8,11,13,14,15
[2]    385     0.9874319     4,5,11,13,14,15
[3]    340     0.9773775     4,11,13,14,15
[4]    248     0.9706745     11,13,14,15
[5]    171     0.9400922     11,14,15
[6]    119     0.8504399     11,14

\end{Verbatim}
For a budget level not in the list, one can look up the model schedule and interpolate the expected predictive accuracy.
For example, for a budget $B \in [171,248)$, the expected predictive accuracy would be 0.9400922 (a conservative estimation).
\begin{figure}
\centering
\begin{center}
\hspace{0cm}
\includegraphics[scale=0.48,clip,angle=-90]{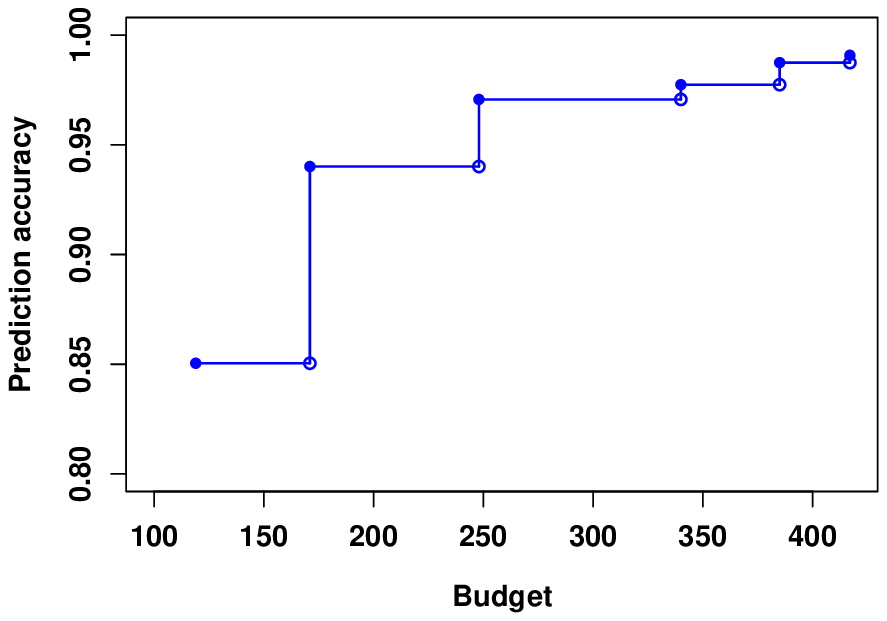}
\end{center}
\abovecaptionskip=-5pt
\caption{\it A graphical representation of the model schedule produced according to the importance of variables. 
Open circles on the curve mean that the corresponding points are not defined according to conventions in mathematics. } 
\label{figure:NPseqI}
\end{figure}
A visualization of the model schedule is shown in Figure~\ref{figure:NPseqI}. The {\it staircase curve} 
shows the expected predictive accuracy at different budget levels. 
\\
\\
A similar model sequence can be generated by using the cost profile of variables. We start with the full model.
Then we recursively prune the most expensive variable that remains until we are left with two variables. Necessary 
bookkeeping allows us to construct a model schedule similarly as that by variable importance. The resulting 
model schedule will be visualized along with that by other model sequences in Figure~\ref{figure:NPindividuals}.
\\
\\
The above two model sequences are generated according to a single metric, the importance or the cost of 
variables. However, the model schedule clearly depends on both, maybe also other factors, in a complicated 
way. As a simple case to start, one can assume that the dependance is only on the importance and the cost, 
and is captured by a function $f(b_i, I_i)$ where $b_i$ and $I_i$ are the cost and importance of the $i\textsuperscript{th}$
variable, such that $f$ is proportional to the variable importance and inversely proportional to its cost. Here 
we consider a simple case 
\begin{equation}
\label{eq:normalizedCost}
f(b_i,I_i) = (I_i/b_i)^{\gamma},
\end{equation}
where $\gamma$ is a parameter (set to be $0.1$ in this work). $f$ is called the {\it normalized importance} of a 
variable. Other choices of $f$ include $f(b_i, I_i)=\alpha_1  I_i + \alpha_2  (1/b_i)$, which we leave 
to future work. We will start from the full model, then sample variables at a probability inversely proportional to 
their normalized importance. Once a variable is selected, it is removed from the current model. That is, less 
`important' variables are removed from the model first. This continues until only two variables 
are left. Related to this, a model sequence can be generated by sampling the variables uniformly at random. 
Note that the sampling procedure introduces randomness in the selection of variables; if the total cost for a 
particular model in the sequence exceeds the given budget, then it would be discarded. Again, the 
resulting model schedule is illustrated in Figure~\ref{figure:NPindividuals}.
\subsubsection{Model sequence by $L_1$-regularization path}
\label{section:methodL1}
The 
$L_1$-regularization path, as a way of generating a model sequence, is attractive for its computational efficiency, 
and efficient algorithms \cite{ParkHastie2007,glmnet2010} have been developed to generate the entire regularization 
path. In this work, we use the {\it glmnet()} package \cite{glmnet2010} for generating the $L_1$-regularization path.
\\
\\
By following the $L_1$ regularization path, i.e., run RF on the set of variables corresponding to a different $\lambda$, 
one can keep track of the model cost (that is, {\it the total cost of all variables with a nonzero coefficient}) for each model 
along the path. The predictive accuracy can be evaluated on a validation set. From models along the regularization 
path, one can produce a model schedule. The user can then pick a model from the schedule with the best accuracy such 
that the total model cost is under a budget $B$, or to trade off between cost and accuracy.  
\\
\\
Next we give an algorithmic description of generating a model sequence by following the $L_1$-regularization path. 
Let $(\bm{X},Y)$ be the input data. Let vector $\bm{B} \in \mathbb{R}^p$ be the cost profile for the $p$ variables. 
Assume there are $J$ classes. Let $n_{\lambda}$ be the number of different $\lambda$'s we take along the regularization 
path. Let $\Theta_{J \times p \times n_{\lambda}}$ store the fitted coefficients for $L_1$-logistic regression, i.e., it consists 
of coefficients for each variable and each class for each step along the regularization path (totally $n_{\lambda}$ steps).
Let $\mathcal{M}_L$ be the model schedule produced by following the $L_1$-regularization path.
The algorithm is described as Algorithm~\ref{algorithm:modelSeqL}. 
\begin{algorithm}{\it~~modelSeqL(X, Y, $\bm{B}$)}
\label{algorithm:modelSeqL}
\begin{algorithmic}[1]
\STATE Invoke $glmnet()$ with the {\it training} data; 
\STATE Initialize the model schedule $\mathcal{M}_L \gets \emptyset$; 
\FOR {$i=1$ to $n_{\lambda}$}
	\STATE Let $\alpha_i$ be predictive accuracy on the validation set; 
	\FOR {$j=1$ to $J$}
		\STATE Let $V_j$ store the index of variables used for class $j$; 
	\ENDFOR
	\STATE Set $V_{used} \gets  \cup_{j=1}^J V_j$; 
	\STATE Calculate total cost $\beta_i$ of all variables based on $V_{used}$ and $\bm{B}$; 
	\STATE Add the new model to model schedule by $\mathcal{M}_L \gets \mathcal{M}_L \cup \{(\beta_i, \alpha_i, V_{used})\}$; 
\ENDFOR
\STATE Return($\mathcal{M}_L$); 
\end{algorithmic}
\end{algorithm} 
Generating a model sequence by following the $L_1$-regularization path involves the computation 
of predictive accuracy by RF on the validation set for each of the $n_{\lambda}$ set of variables, the 
calculation of the total variable costs and necessary bookkeeping along the regularization path. 
In total this costs $O(n_{\lambda} \cdot (T \cdot p^{1/2} \cdot n(\log(n))^2) )$. Thus the computational 
complexity for Algorithm~\ref{algorithm:modelSeqL} is 
\begin{eqnarray*}
&& O\left(n_{\lambda} \cdot p\cdot J\cdot E\cdot n + n_{\lambda} \cdot (T \cdot p^{1/2} \cdot n(\log(n))^2 ) \right) \\
&=& O\left(n_{\lambda} \cdot max\left(p^{1/2} \cdot J\cdot E, ~T \cdot (\log(n))^2\right) \cdot p^{1/2} \cdot n \right).
\end{eqnarray*}
It follows that the overall computational complexity of our algorithm is give by 
$O\left(n_{\lambda} \cdot max\left(J\cdot E, ~T \cdot p^{1/2}\cdot (\log(n))^2\right) \cdot p \cdot n \right)$.
\subsubsection{Example model schedules for the NPP dataset}
\label{section:methodExamples}
\begin{figure}[htbp]
\centering
\begin{center}
\hspace{0cm}
\includegraphics[scale=0.5,clip,angle=-90]{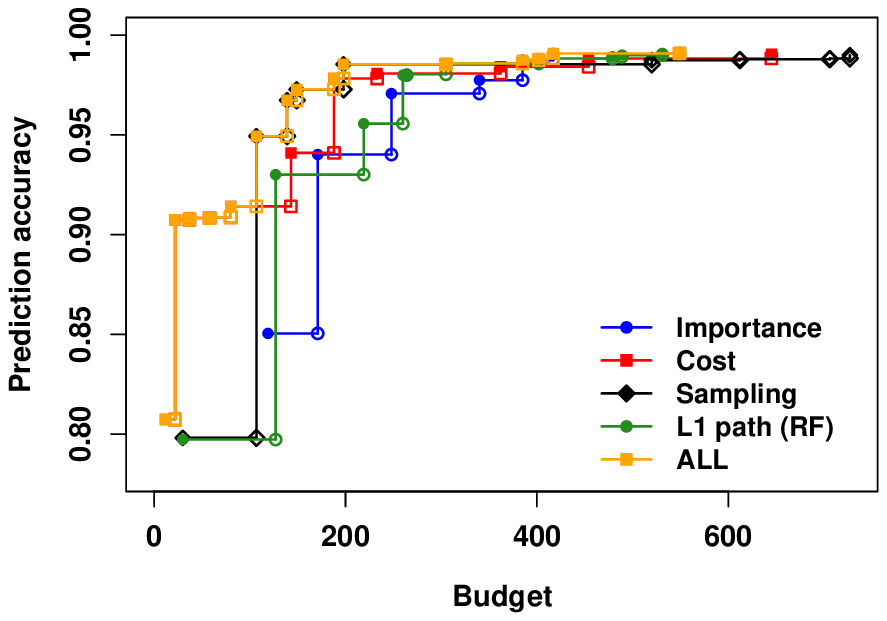}
\end{center}
\abovecaptionskip=-5pt
\caption{\it The ensemble (indicated by `All') vs individual model schedules. } 
\label{figure:NPensemble}
\end{figure}

Figure~\ref{figure:NPindividuals} shows the model schedule produced by four different ways of generating model 
sequences---by variable importance, variable cost, sampling with normalized variable importance, and $L_1$-regularization 
path. It can be seen that each resulting model schedule has its own merit, and no one dominates 
others. By `dominate' we mean the staircase curve corresponding to one model schedule is higher than that 
of another at {\it all} different budget levels.
\begin{figure}[htbp]
\centering
\begin{center}
\hspace{0cm}
\includegraphics[scale=0.5,clip,angle=-90]{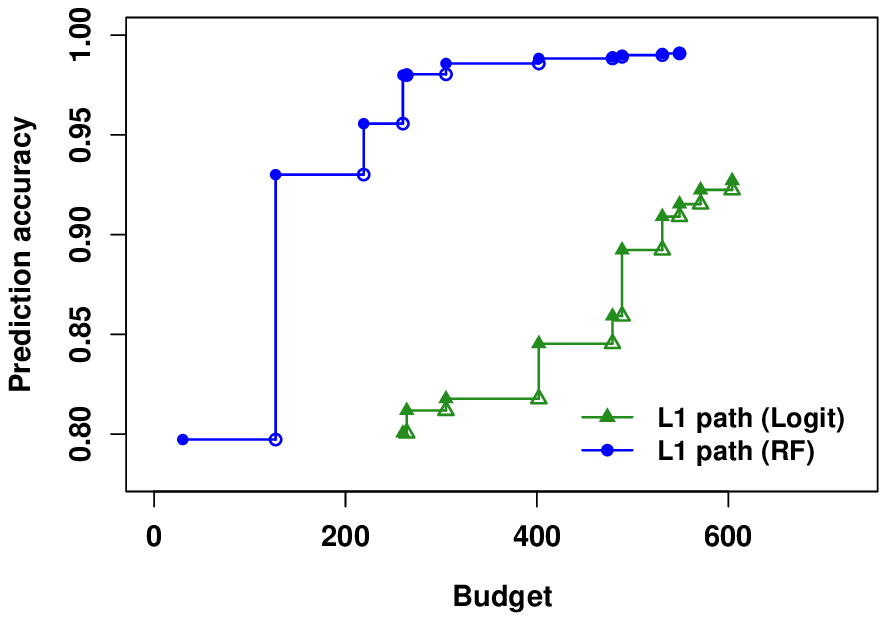}
\end{center}
\abovecaptionskip=-5pt
\caption{\it RF Vs logistic regression by following the same $L_1$-regularization path.  } 
\label{figure:NPL1RF2LR}
\end{figure}
\\
\\
A curious question is, does RF improve over $L_1$ logistic regression, if following the same $L_1$-regularization
path? The answer is `Yes' for the NPP dataset. This is illustrated in Figure~\ref{figure:NPL1RF2LR}, which shows that the 
model schedule produced by RF dominates that by $L_1$ logistic regression with a large margin. This gives support 
for RF to be a preferred engine for the selection and combination of variables in some applications.
\\
\\
A more important question is, does ensemble, i.e., a model schedule produced by combining those generated by different 
ways, improve the model schedule? The answer is `Yes' for the NPP dataset. Figure~\ref{figure:NPensemble} is an 
illustration. The staircase curve by an ensemble of four model sequences dominates those by any individual ones. Indeed 
this is a consequence of the way that different model schedules are combined in our approach, and by packing more model
sequences into the ensemble will results in a better model schedule.
\subsection{Algorithmic description}
\label{section:methodAlg}
In this section, we will describe algorithms for variable selection with {\it msB}. Let $\mathcal{M}$ be the {\it final model schedule} 
produced by {\it msB}. That is, by combining model schedules generated by members in the ensemble; our current 
implementation consists of model schedules generated by variable cost, by variable importance, by sampling, and by 
following $L_1$-regularization path. The combining of multiple model sequences is implemented by treating each model 
sequence as a {\it set} of triples {\it (accuracy, cost, variables)}, then we take the union of all such sets from individual model 
sequences. This is then compressed by removing those triples corresponding to a higher cost but lower predictive accuracy. 
Our approach is implemented as three algorithms
\begin{itemize}
\item  $msB()$ for generating the final model schedule from multiple model sequences
\item $modelSeq()$ for generating a model schedule for a given type of model sequence
\item $modelSeqL()$ for generating a model schedule by following the regularization path of $L_1$ logistic regression.
\end{itemize}
The algorithm for $modelSeqL()$ is described as Algorithm~\ref{algorithm:modelSeqL} in Section~\ref{section:methodL1}, and
that for $msB()$ and $modelSeq()$ are described as Algorithm~\ref{algorithm:msB} and Algorithm~\ref{algorithm:modelSeq}, 
respectively, in the rest of this section, where $\bm{W}_i$ and $\bm{W}_u$ denote the importance and normalized 
importance of variables, respectively. 
\\
\begin{algorithm}{\it~~msB(X, Y, $\bm{B}$)}
\label{algorithm:msB}
\begin{algorithmic}[1]
\STATE Invoke RF with the training and validation set; 
\STATE Generate model sequence by variable cost $\mathcal{M}_C \leftarrow modelSeq(X,Y,\bm{B},2)$; 
\STATE Generate model sequence by variable importance $\mathcal{M}_I \leftarrow modelSeq(X,Y,\bm{B},1)$; 
\STATE Generate model sequence by sampling $\mathcal{M}_S \leftarrow modelSeq(X,Y,\bm{B},3)$; 
\STATE Generate model sequence by $L_1$-logit $\mathcal{M}_L \leftarrow modelSeqL(X,Y,\bm{B})$; 
\STATE Combine model sequences by set union $\mathcal{M} \leftarrow \mathcal{M}_C \cup \mathcal{M}_I \cup \mathcal{M}_S \cup \mathcal{M}_L$; 
\STATE Sort model list $\mathcal{M}$ by model cost associated with each entry in the list; 
\STATE For each model cost in $\mathcal{M}$, delete $\mathcal{M}$'s entries with higher cost but lower accuracy;
\STATE Return($\mathcal{M}$); 
\end{algorithmic}
\end{algorithm} 
\begin{algorithm}{\it~~modelSeq(X, Y, $\bm{B}$, type)}
\label{algorithm:modelSeq}
\begin{algorithmic}[1]
\STATE Invoke RF with the training and validation set; 
\STATE Let $\alpha_0$ be the predictive accuracy on the validation set; 
\STATE Set variable cost for full model $\beta_0 \leftarrow \sum_{i=1}^p b_i$; 
\STATE Set variable weights $w \leftarrow \bm{W}_i$ if $type==1$; 
\STATE Set variable weights $w \leftarrow 1/\bm{B}$ if $type==2$; 
\STATE Set variable weights $w \leftarrow \bm{W}_u$ if $type==3$; 
\STATE Initialize the list of models $\mathcal{M}  \leftarrow (\beta_0, \alpha_0, V)$; 
\FOR {$i=1$ to $p-2$}
	\IF{$type == 3$}
		\STATE Sample variable $v \in V$ with weights $\bm{W}_u$ and set $V \leftarrow V-\{v\}$;
	\ELSE
		\STATE Let $v  \leftarrow \arg \min_{i \in V} \{w[i]: ~ i \in V\}$ and set $V \leftarrow V-\{v\}$;
	\ENDIF;
	\STATE Invoke RF on variables from the set $V$; 
	\STATE Let $\alpha_{V}$ and $\beta_V$ be the predictive accuracy and model cost; 
	\STATE Append the new model by set union $\mathcal{M} \leftarrow \mathcal{M} \cup \{(\beta_V, \alpha_V, V)\}$; 
\ENDFOR
\STATE Return($\mathcal{M}$); 
\end{algorithmic}
\end{algorithm} 
Note that our approach does not exclude the use of classifiers other than RF, neither does 
it have a restriction on the number of model sequences, in the ensemble. 
\\
\\
The computational complexity of producing a model sequence by variable importance is calculated as follows. 
The generation of variable importance by RF costs $O(T \cdot p^{3/2} \cdot n(\log(n))^2)$ (the ranking of variable
importance is absorbed in this term), while that by normalized importance is $O(p)$. Then there is a sequential 
removal of variables with each round involving the selection or sampling of variables which costs $O(p)$, and 
the fitting of RF and then applying to the validation set which costs $O(T \cdot p^{1/2} \cdot n(\log(n))^2 )$, so this step
costs $O(T \cdot p^{3/2} \cdot n(\log(n))^2)$ in total. Thus the computational cost for the generation of each of these three 
model sequences, or each invocation of Algorithm~\ref{algorithm:modelSeq},  is $O(T \cdot p^{3/2} \cdot n(\log(n))^2)$. 
Algorithm 3.2 is an invocation of Algorithm 3.1 and that of Algorithm 3.3 for 3 times plus some post-processing, thus its computational 
complexity is $O\left(n_{\lambda} \cdot max\left(J\cdot E, ~T \cdot p^{1/2}\cdot (\log(n))^2\right) \cdot p \cdot n \right)$.
Putting the computational complexity of individual algorithms together gives the overall computational complexity of our algorithm 
as $O\left(n_{\lambda} \cdot max\left(J\cdot E, ~T \cdot p^{1/2}\cdot (\log(n))^2\right) \cdot p \cdot n \right)$. It scales linearly, with 
some additional $\log$-factors, with the the number of data instances $n$.
\section{Optimality and a toy example}
\label{section:optimality}
Just like any machine learning problem, it is always important to consider optimality---what would be the `optimal' model 
schedule for a given data distribution and cost profile? Let us focus on the classification problem. In the following, we will 
define the optimal model schedule, and give a toy example to demonstrate that our algorithm can deliver a near optimal 
model schedule.
\\
\\
Assume the data is generated from a distribution
(often unknown) in $\mathbb{R}^p \otimes \mathcal{J}$ where $\mathcal{J}=\{1,2,...,J\}$ is the set of labels. Assume the 
cost profile of variables is given by $\bm{B}$. 
Let $V=\{1,2,...,p\}$ be the set of indices of all the $p$ variables. Then the set of all possible combinations of variables is 
given by $\mathcal{V}=\{S:~ S\subseteq V ~\mbox{s.t.}~ |S|>0\}$. Let $g(S)$ denote the total cost of variables in set $S$. 
For a given data, the set $\mathcal{V}$ is finite. Thus, there are only finite possible values for the total cost of variables used 
in the model; let $\mathcal{C}$ denote the set of possible costs. Then, for a given data distribution, the {\it optimal model schedule} 
is defined as the following collection of pairs (here for simplicity we omit such information as variables used, coefficients 
etc in the model schedule)
\begin{equation*}
\left\{(c,\alpha):~ c\in \mathcal{C}, \alpha=\max_{S \in \mathcal{V}, g(S) =c} \left\{\mbox{Bayes rate on feature set S} \right\} \right\}.
\end{equation*}
The above defines the best predictive accuracy for each possible cost level.
If a universally consistent classifier, such as AdaBoost with early stop \cite{BartlettTraskin2007}, support vector machines 
\cite{CortesVapnik1995} etc is used, then the Bayes rate can be achieved on any subset of variables as long as such a 
subset is visited by the algorithm. Thus, it is desirable to include a universally consistent classifier in the algorithm (RF is 
used in our algorithm due to its superior empirical performance though its universal consistency is still unknown), and then 
the remaining issue is to try to hit as many critical points in the solution space as possible. That is the idea of our approach. 
We will use a toy example to illustrate this.  
\\
\\
The toy example we choose is a small scale problem for which the optimal model schedule can be computed by exhaustive
search. The data is generated by a 4-component Gaussian mixture in $\mathbb{R}^8$ specified as 
\begin{equation*}
\frac{1}{4}\sum_{i=1}^4 \mathcal{N}(\mu_i, \Sigma), 
\end{equation*}
where the covariance matrix $\Sigma$ is defined by 
\begin{equation*}
\Sigma_{i,j}=\rho^{\vert i-j \vert}, ~~\mbox{for}~ \rho=0.1, 0.3, 0.6,
\end{equation*}
and the center of the four components are 
\begin{eqnarray*}
&& \mu_1=(2.0,1.8,1.6...,0.6), ~\mu_4=-\mu_1,
\end{eqnarray*}
with $\mu_2$ and $\mu_3$ the same as $\mu_1$ except that the second half and the first half of their components are taking 
an opposite sign. To introduce variety into the underlying data, we let $\rho$ vary over $\{0.1,0.3,0.6\}$. The mixture ID is used
as the label for each data instance. The sample size is 50,000 with 60\% used for training, 20\% for the selection of models in 
individual model sequences (validation set), and 20\% for producing the predictive accuracy by the final model schedule. 
The sample size is chosen to be large enough so that the predictive error rate stops decreasing when further increasing the sample 
size. The cost of variables are set as follows (produced by sampling from $[1,100]$ uniformly at random and then stay fixed)
\begin{equation*}
92, 81, 45, 23, 23, 33, 72, 5.
\end{equation*}   
Since there are only 8 variables for this classification problem, we can try all possible (totally 255) combinations of variables.
For each combination, the total cost of all involving variables is calculated, and the predictive accuracy by RF is assessed.
The optimal model is found by an exhaustive search over all combinations of variables. A similar approach was taken by \cite{MinHQZ2011}.
\begin{figure}
\centering
\begin{center}
\hspace{0cm}
\includegraphics[scale=0.65,clip,angle=-90]{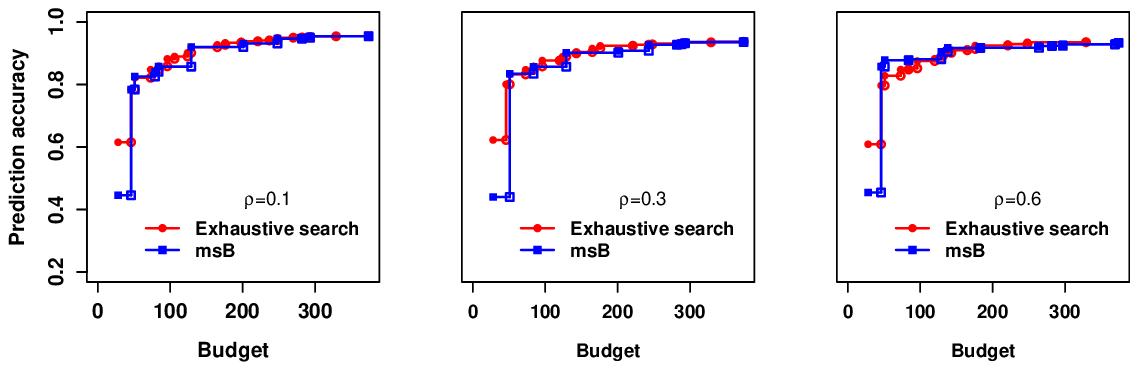}
\end{center}
\abovecaptionskip=-5pt
\caption{\it Model schedule produced by exhaustive search and by {\it msB} (our approach) on the Gaussian mixture data. } 
\label{figure:exMsB}
\end{figure}
\\
\\
Figure~\ref{figure:exMsB} shows the model schedule found by exhaustive search and by our algorithm, under different
values of $\rho$. In all cases, the model schedules found by the two are nearly identical. 
For this problem, the total number of variable combinations, or candidate pairs $(cost, accuracy)$, is about 250 (excluding 
cases with only one variables for which RF cannot run). We term the collection of all candidate pairs as the {\it solution 
space} or model space. Our algorithm only visits a small fraction, about 30/250=12\%, of the solution space, but does 
surprisingly well in producing the model schedule. 
\begin{figure}[htbp]
\centering
\begin{center}
\includegraphics[scale=0.5,clip,angle=-90]{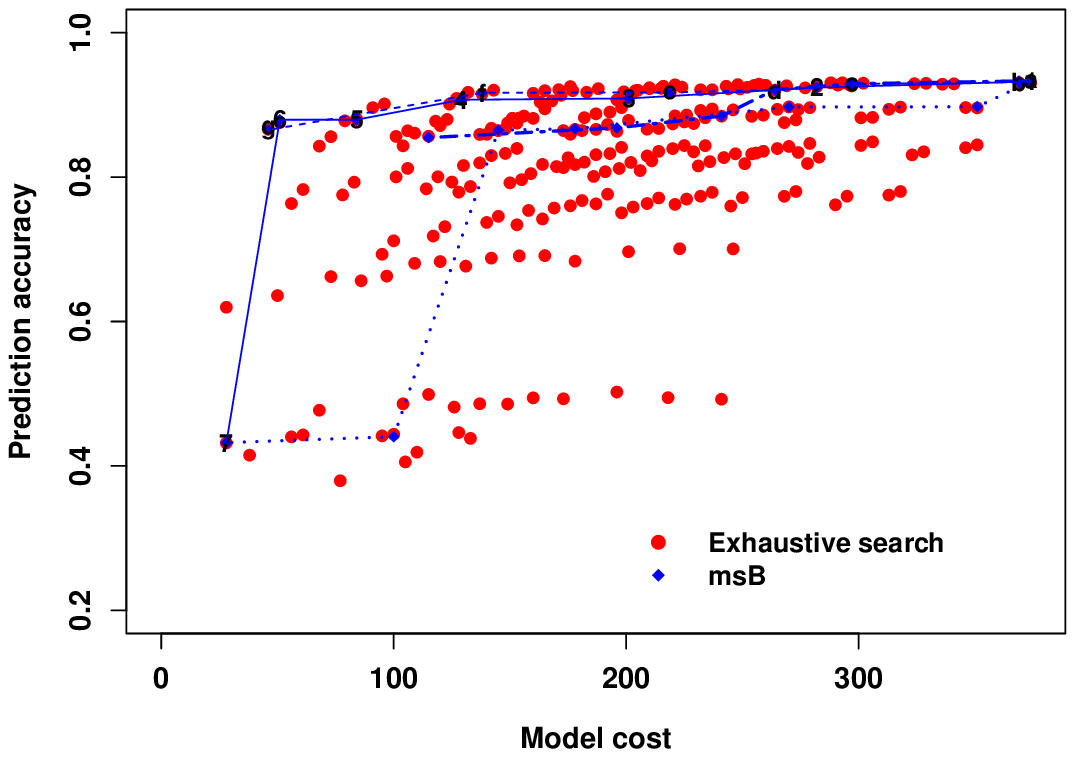}
\end{center}
\abovecaptionskip=-5pt
\caption{\it Points in the solution space that are visited by exhaustive search and by {\it msB}. For points in the solution 
space visited by {\it msB}, there are 4 sequences, marked by numbers "1-2-3-4-5-6-7", letters ``a-b-c-d-e-f-g", by dotted line, 
and by two-dash line, corresponding to variable selection by cost, accuracy, sampling, and $L_1$-regularization path, 
respectively. } 
\label{figure:gmix6SolPts}
\end{figure}
To uncover the mystery, we plot the solution space, and then mark points visited by our algorithm in Figure~\ref{figure:gmix6SolPts}. 
Our algorithm is very efficient in that it visits only the most {\it promising} part, a small fraction, of the solution space. 
In particular, the two sequences of points, marked as ``1-2-3-4-5-6-7" and ``a-b-c-d-e-f-g" (produced by selecting 
variables according to variable cost and variable importance, respectively), almost always stay close to the `optimal' 
part of the solution space.
\\
\\
The reason why our algorithm is efficient in finding `promising' search paths can be understood as follows. Starting 
at the full model (top right corner in Figure~\ref{figure:gmix6SolPts}), our algorithm successively removes the most 
expensive variable (by cost) or the weakest variable (by importance), this effectively does a
gradient descent in some functional space, i.e., follows the direction along which the model cost decreases the most (i.e., 
a cost-greedy direction) or accuracy decreases the least, thus the next point visited by our algorithm in the model space 
will be either a point that is cheaper in model cost but with potentially similar (maybe even better) 
predictive accuracy (since the weakest variable is removed), or much cheaper (since the most expensive variable 
is removed) in cost but potentially not much reduction in accuracy. Such moves in the solution space 
are desirable in reaching an economic model schedule. Of course, other model sequences adopted by our algorithm may 
potentially correct sub-optimal moves, or throw in some better moves along the way. The overall effect is, by visiting a 
small part of the solution space we have already seen the `best' part of the world.  
\\
\\
Some additional insights can be gained by comparing to algorithms trained on a randomly selected sizable fraction, such as 1/2, 1/3 etc, 
of the variables. Assuming that the variables satisfy some kind of redundancy and that the number of pure noise variables is `small', then
with high probability the loss in Bayes rate for classification is negligible when a random selection of a sizable fraction of variables is used \cite{TACOMA2012}. 
It is sensible to assume that by ensemble of a number of search paths would do better that random selection, thus 
we can expect that the loss in classification accuracy (compared to the Bayes rate) would be small by our algorithm on large training data.
\section{Experiments}
\label{section:experiments}
We conduct a range of experiments on several different aspects. We compare the model schedules generated by {\it msB}, 
and a competing method \cite{L1Budget2018}, denoted by {\it logitB}, as that is the only 
method available that strictly enforces the budget constraint. 
The comparison is done both visually and numerically under a quantitative metric. We also compare our approach to 
the method proposed in \cite{ZhouZhouLi2016}, denoted by {\it rfB}. As {\it rfB} does not produce a 
model schedule, we obtain the total cost, $\nu$, incurred by {\it rfB} first, 
and then compute the predictive accuracy achieved by {\it msB} under budget $\nu$. The computation 
time is also compared among different methods. We also explore the feasibility of exhaustive search as well as the 
optimality of our approach by comparing our approach to exhaustive search. We report the results in Section~\ref{section:expLogitB}, Section~\ref{section:expRfB}, and Section~\ref{section:expXsearch}, respectively. We start by an introduction of 
the datasets used in our experiments.
\subsection{Datasets}
\label{section:datasets}
We conduct experiments on a number of datasets, including nine from the UC Irvine Machine Learning Repository 
\cite{UCI} and an additional remote sensing data adopted from a recent study \cite{rsDiagnosis2019}. The
UC Irvine datasets are the {\it Naval propulsion plants (NPP), Steel plate faults, Spam filter, Concrete compressive 
strength, Landsat, Thyroid disease, Vehicle silhouette, Bank marketing}, and {\it US census income (USCI)}. A summary 
of the datasets is given in Table~\ref{table:datasets}.
\\
\\
Two of the UC Irvine datasets were originally used for regression and we convert the relevant output variable(s) so 
that they could be used for classification. These are the Naval propulsion plants data and the Concrete compressive 
strength data. For the former, we treat any record of measurements as requiring maintenance if both the {\it q3Compressor} 
and {\it q3Turbine} variables are above their median values. For the latter, we convert the data by rounding the compressive 
strength variable to 4 categories according to its 4 quartiles. The original USCI data has 299,285 instances with 41 
features. We follow the preprocessing procedure in \cite{distSpectArXiv2019}, and exclude instances with missing 
values, and also features \#26, \#27, \#28 and \#30, due to too many values. This leaves 285,799 instances on 37 
features, with all categorical variables converted to integers. For the Bank marketing data, we follow \cite{distStatArXiv2019} 
where all features are converted into numeric values and then standardize.     
\\
\\
The remote sensing data is about a region, spanning 23\degree 2'-23\degree 25'N, 113\degree 8'-113\degree 35'E, 
in Guangdong Province of South China. There are 7 different land-use types (classes), including {\it water, residential 
area, natural forest, orchard, industry or commercial area, idle land}, and {\it bareland}. The features were derived from 
a Landsat Thematic Mapper (TM) image about the region acquired in January 2009. There are totally 56 features, including 
6 spectral features corresponding to the 6 TM bands, 8 texture features, {\it mean, variance, homogeneity, contrast, dissimilarity, 
entropy, second moment, correlation}, for each of the 6 TM bands, and two location features, the {\it latitude} and {\it longitude} 
of the ground position associated with each data instance. 
\begin{table}
\begin{center}
\begin{tabular}{r|rrr}
\hline
   \bf{Dataset}                    				& \bf{Features}     	& \bf{Classes}   	&  \bf{Instances}\\[1pt]
\hline \\[-10pt]
Naval propulsion plants  				&16   	     			&2         			&11934\\
Steel plate faults        					& 27                  	&7           			& 1941\\
Spam filter                         					& 57                  	&2           			& 4601\\
Concrete strength       				&8					&4					&1030\\
Remote sensing                     & 56                  	&7           			& 3303\\
Landsat                         				& 36                  	&6           			& 4435\\
Thyroid disease						&21				&3					&7200\\
Vehicle silhouette						& 18					&4					&846\\[1pt]
Bank marketing						& 16					&2					&45211\\[1pt]
US censors income					& 40					&2					&285779\\[1pt]
\hline
\end{tabular}
\caption{A summary of datasets used in experiments.} 
\label{table:datasets}
\end{center}
\end{table}
\subsection{Comparison with {\it logitB}}
\label{section:expLogitB}
\begin{figure}
\centering
\begin{center}
\hspace{0cm}
\includegraphics[scale=0.39,clip,angle=-90]{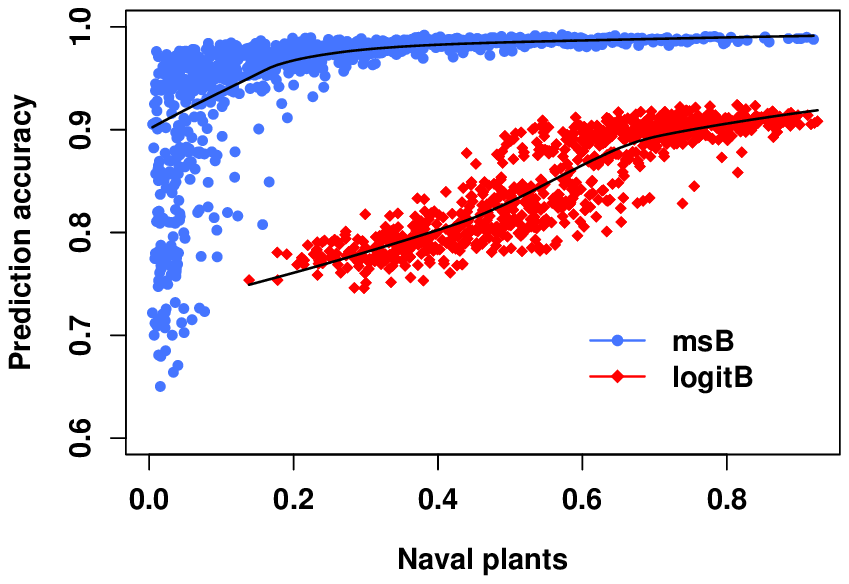}
\includegraphics[scale=0.39,clip,angle=-90]{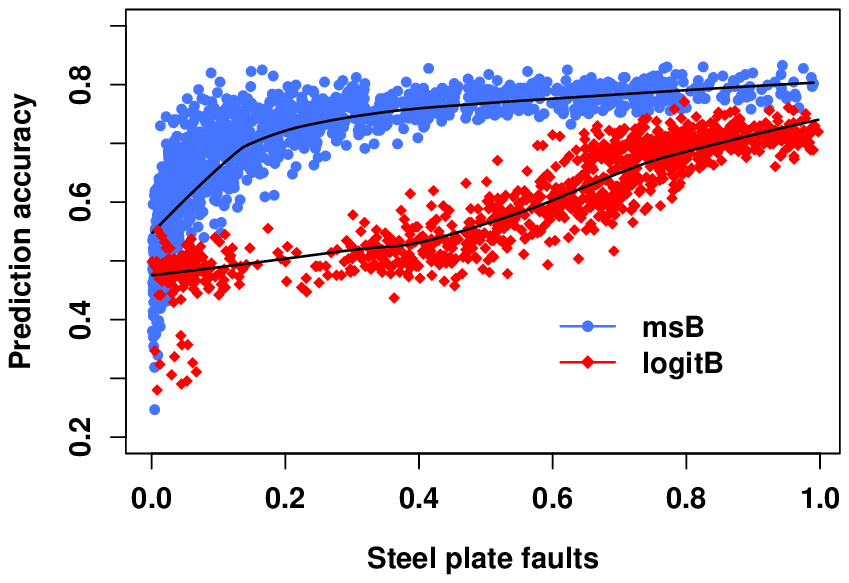}
\includegraphics[scale=0.39,clip,angle=-90]{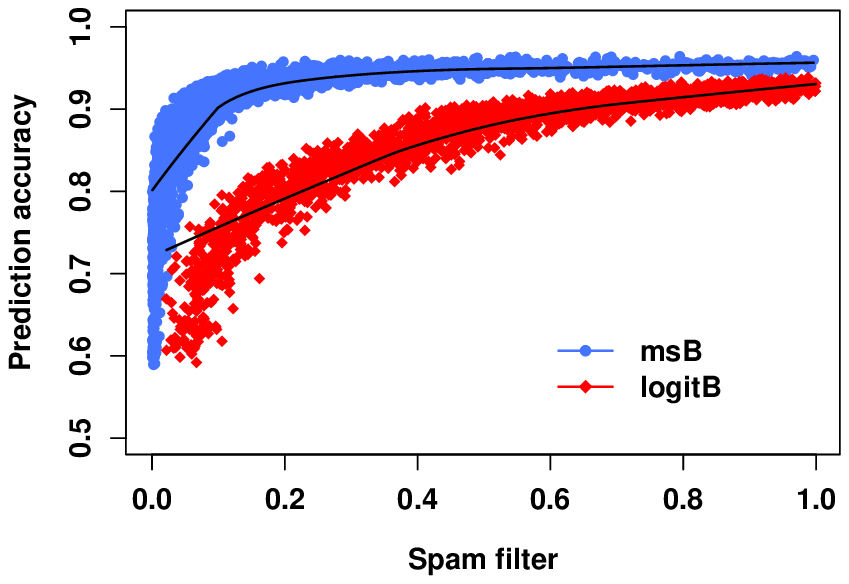}
\includegraphics[scale=0.39,clip,angle=-90]{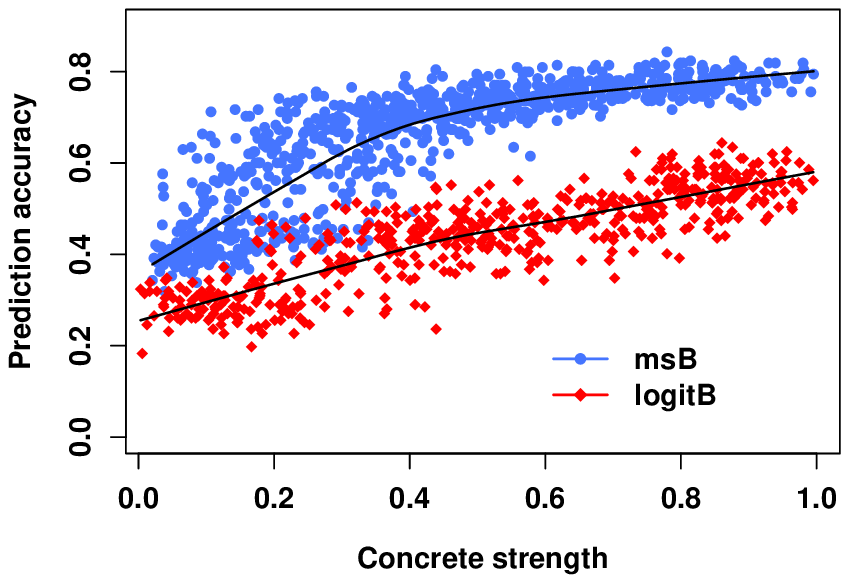}
\includegraphics[scale=0.39,clip,angle=-90]{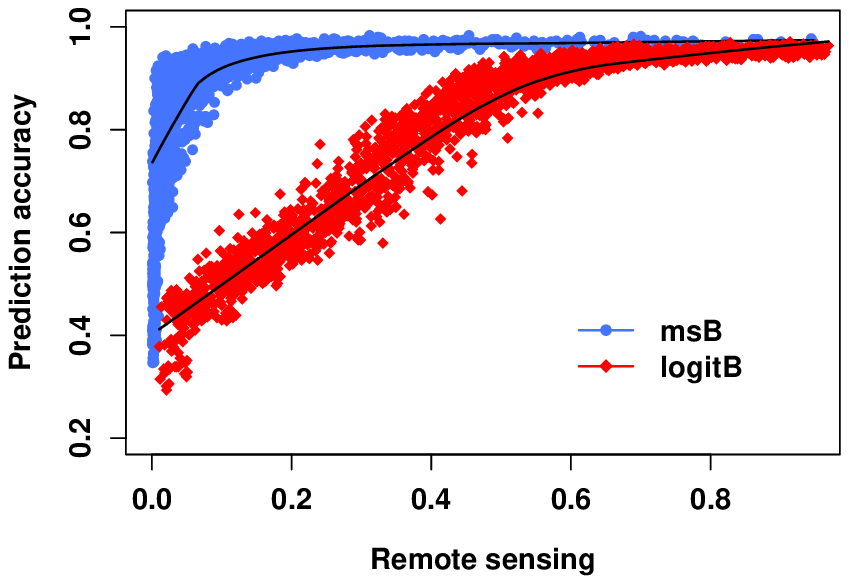}
\includegraphics[scale=0.39,clip,angle=-90]{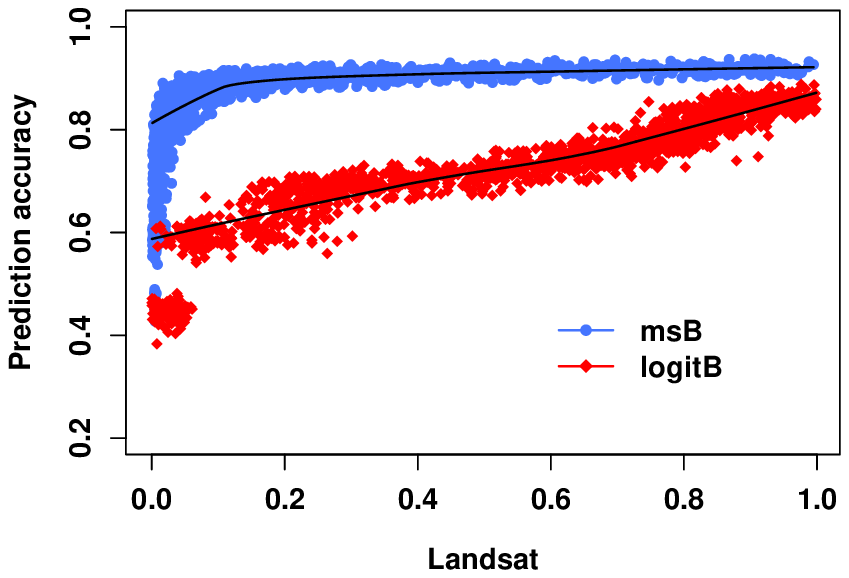}
\includegraphics[scale=0.39,clip,angle=-90]{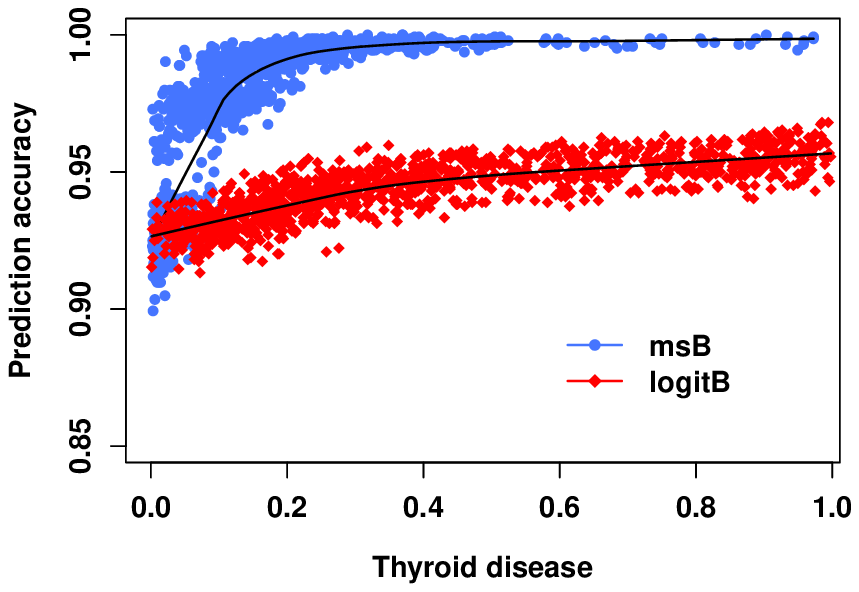}
\includegraphics[scale=0.39,clip,angle=-90]{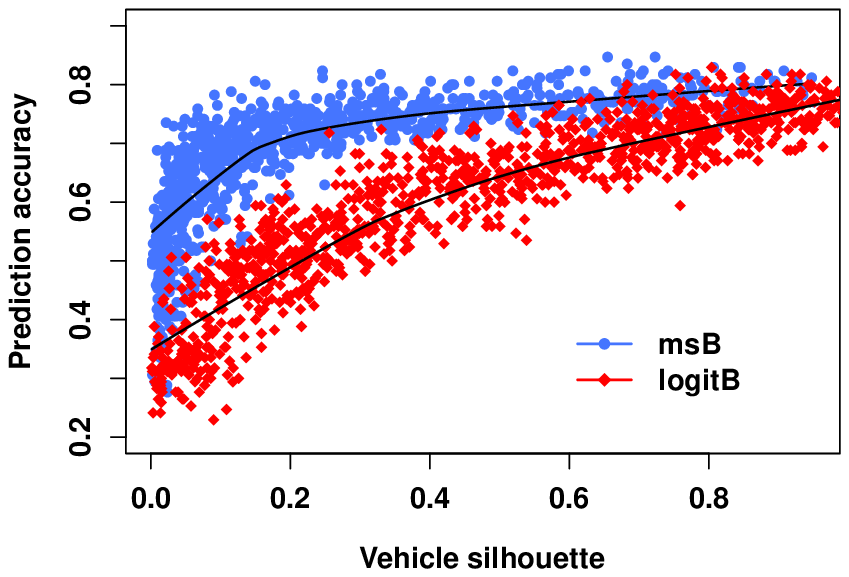}
\end{center}
\caption{\it The expected predictive accuracy by {\it logitB} and {\it msB} for 8 datasets used in our experiment. 
The x-axis indicates the normalized model cost, which is the relative cost with respect to the cost for the full model. 
{\it msB} and {\it logitB} indicate our approach and $L_1$ logistic regression under a budget, respectively.} 
\label{figure:curves8data}
\end{figure}
\begin{figure}
\centering
\begin{center}
\hspace{0cm}
\includegraphics[scale=0.35,clip]{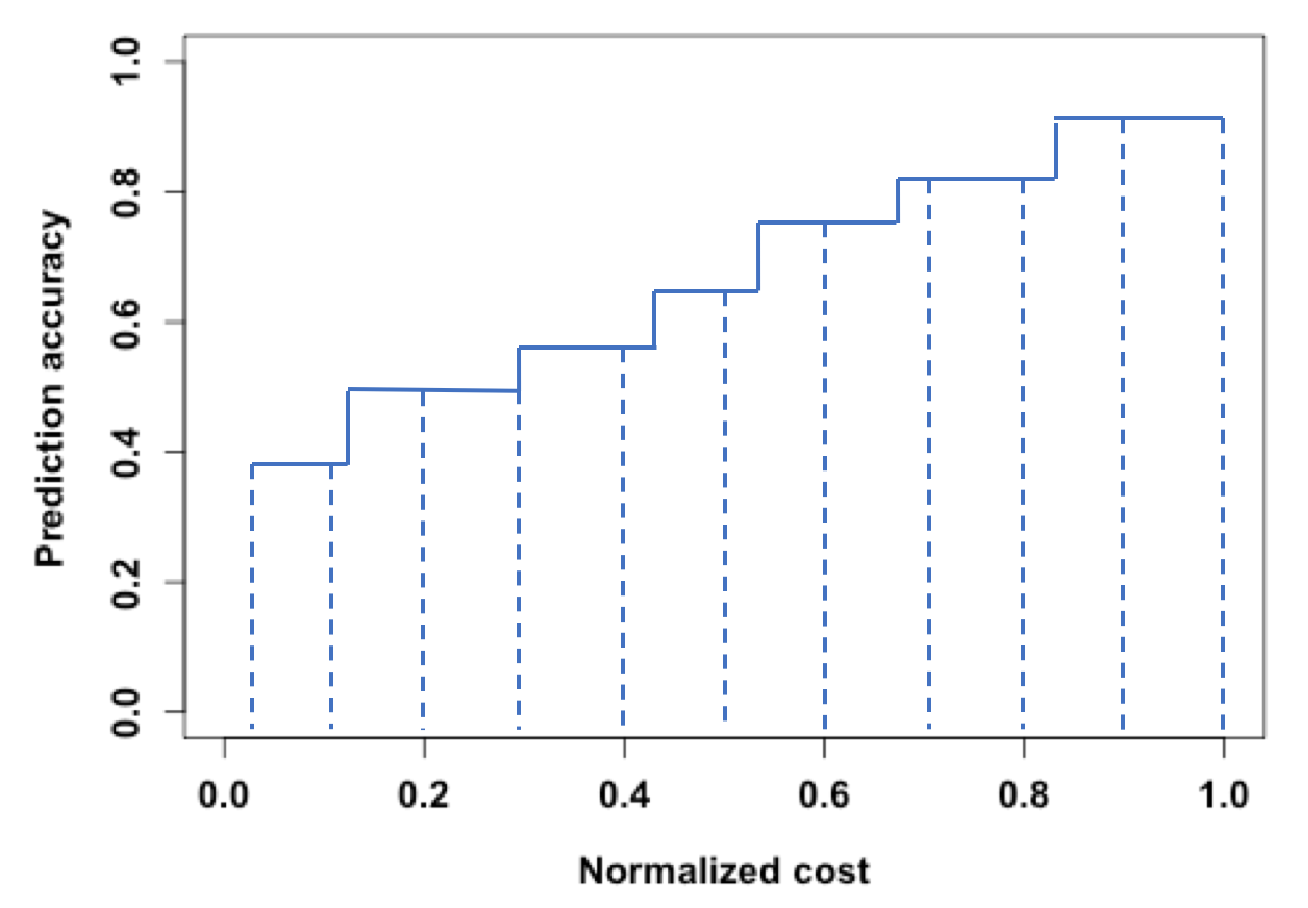}
\end{center}
\abovecaptionskip=-5pt
\caption{\it Illustration of the area under the Pareto (dashed area in the figure).  } 
\label{figure:AUSIllus}
\end{figure}
\begin{figure}[htbp]
\centering
\begin{center}
\hspace{0cm}
\includegraphics[scale=0.56,clip,angle=-90]{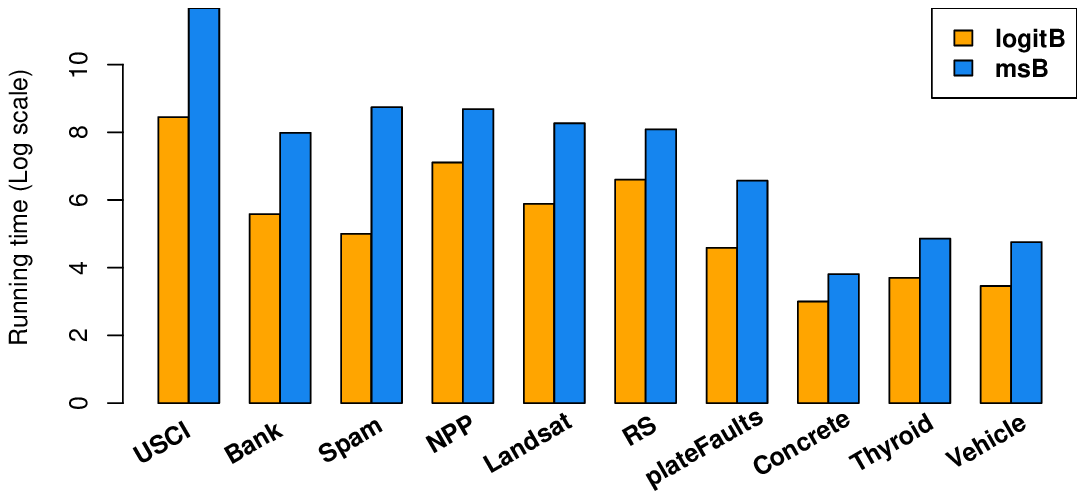}
\end{center}
\abovecaptionskip=-5pt
\caption{\it A comparison of the running time (in $\log_2$seconds) for model schedules generation by {\it LogitB} and {\it msB}. } 
\label{figure:runningTime}
\end{figure}

The experiments are conducted as follows. For all 
datasets used, a random sample of 60\% of the data are used for training, 20\% for the selection of models in individual 
model sequence, and 20\% for producing the predictive accuracy by the final model schedule. As no 
dataset used in our experiments comes with a cost profile for its variables, we randomly generate the cost 
by sampling from $[1,100]$ uniformly at random. Then a model schedule is generated by {\it msB} and by {\it logitB}. 
This is repeated for 100 runs. As different runs of our experiments are under a different variable cost profile, we 
normalize the model cost by dividing it by the cost of a full model in the same run. 
An {\it average model schedule} is produced by curve smoothing (with {\it lowess()} function in R) over model 
schedules generated over the 100 runs. 
\\
\\
Our final output over 100 runs is visualized as follows. Instead of using the staircase curve (which would make the 
plot too crowded for 100 model schedules), we plot individual pairs, {\it (model cost, predictive accuracy)}, in a model 
schedule as points in a scatter plot. Then we add the {\it average model schedule} (indicated by solid curves) to the 
scatter plot. Figure~\ref{figure:curves8data} shows model schedules generated for the first 8 datasets used in our 
experiment (the other two omitted due to page size limit). It can be seen that, in almost all cases, scatter points 
generated by our algorithm lie substantially higher above those by {\it logitB}. This indicates that, for the same 
normalized model cost, our algorithm could produce a model with substantially higher predictive accuracy. Similarly, 
the average model schedules produced by our algorithm dominate that by {\it logitB} by a substantial margin on all 
datasets used in our experiment. 
\\
\\
Beyond visualization, we also produce a single number metric for the cost-accuracy curve for a given model schedule
similarly as the AUC (area under the curve) \cite{Spackman1989}. This is done by 
scaling the model cost as a proportion of that under the full model, which turns 
the model cost into a number in the range of $(0,1]$. Thus, a model schedule curve, after cost normalization, would 
lie within the unit square defined by $(0,1] \times (0,1]$.
By calculating the area above the x-axis but under the model schedule curve, we obtain a number in the range of $(0,1]$. 
This would be a sensible metric to compare different model schedules, since at any cost level the higher the curve lies, 
the better the accuracy. We call this metric the {\it area under the Pareto} (AUP), as the model schedule curve takes 
the shape of a (irregular) staircase or Pareto fronts. Note that we do not have to restrict to the staircase-shaped curve; the curve can 
be a generic one, for instance the smoothed average model schedule curve over many runs. Figure~\ref{figure:AUSIllus} 
illustrates the area under a model schedule curve, where we are interested in the size of the shaded area. 
\begin{figure}
\centering
\begin{center}
\hspace{0cm}
\includegraphics[scale=0.56,clip,angle=-90]{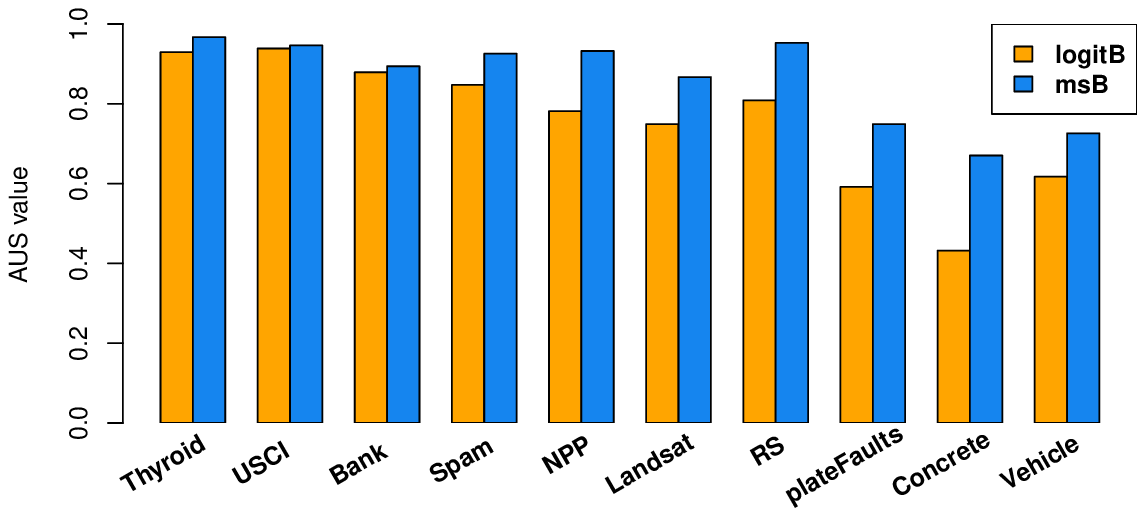}
\end{center}
\abovecaptionskip=-5pt
\caption{\it A comparison of the AUP score for model schedules generated by {\it logitB} and {\it msB}. } 
\label{figure:compAUP}
\end{figure}
\\
\\
Figure~\ref{figure:compAUP} shows the AUP score computed for all the datasets under {\it logitB} and {\it msB}, 
respectively. We can see that {\it msB} outperforms {\it logitB} by a large margin on most of the 10 datasets. We 
also evaluate the running time for {\it logitB} and {\it msB}. The results are shown in Figure~\ref{figure:runningTime}, 
and we see that the {\it logitB} algorithms runs faster than {\it msB} on all the datasets. This is expected as {\it logitB}
only does part of the work of that of {\it msB}.
\subsection{Comparison with {\it rfB}}
\label{section:expRfB}
\begin{figure}
\centering
\begin{center}
\hspace{0cm}
\includegraphics[scale=0.55,clip, angle=-90]{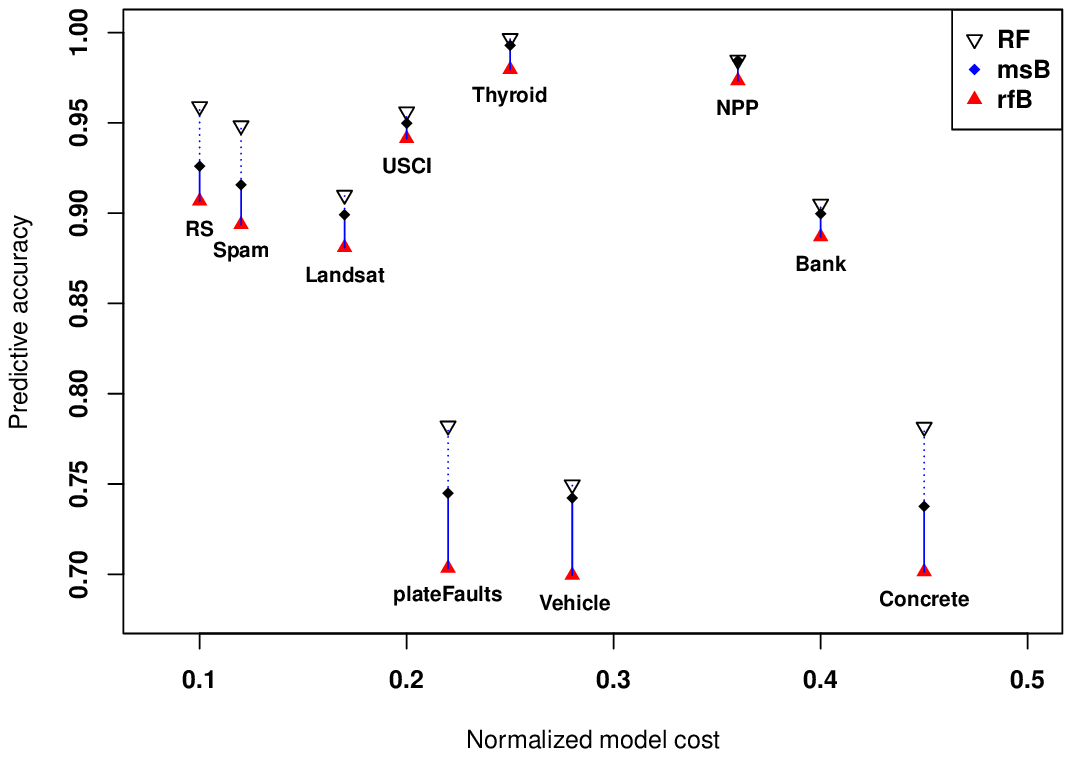}
\end{center}
\abovecaptionskip=-5pt
\caption{\it Comparison of {\it msB} to {\it rfB}. The predictive accuracy by RF is included to indicate what one would expect 
without any budget constraint. } 
\label{figure:Competitor}
\end{figure}

The comparison to another competing algorithm, {\it rfB}, is different. As {\it rfB} does not produce a model schedule, 
we choose to compare its predictive accuracy to that by {\it msB} under similar variable costs. For each dataset, we 
obtain the total cost, $\nu$, incurred by {\it rfB} first, and then we compute the predictive accuracy achieved by {\it msB} 
under budget $\nu$; the respective predictive accuracies are compared. The results are shown in 
Figure~\ref{figure:Competitor}. It can be seen that in most of the cases, {\it rfB} gives inferior predictive accuracy 
than {\it msB} under a similar cost. A limitation with {\it rfB} is that there is neither a guarantee on the total variable cost nor 
that on the predictive accuracy. While the resulting model by {\it rfB} may have a small variable cost, but that is not 
necessary under the predefined budget. This is because while it aims to produce a low-cost model, it 
achieves this by sampling, and thus the variables enter the model partially by chance. When the predefined budget is high, {\it rfB} 
cannot adapt to the high budget to deliver a better predictive accuracy; this can be seen from the often substantial performance 
gap towards the original RF. As {\it rfB} tries to give more weights to low-cost variables, rather than the most predictive
ones, that hurts the predictive accuracy. Thus even at the same cost, the predictive accuracy it achieves may be 
substantially lower than that by {\it msB}. This can be seen from the noticeable gaps in predictive accuracy between 
{\it rfB} and {\it msB}. On the other hand, in several cases, even at a low budget level, {\it msB} is able to achieve the 
level of predictive accuracy by RF under no budget constraint.
\subsection{Feasibility of exhaustive search}
\label{section:expXsearch}
We also explore the feasibility of exhaustive search on three datasets with relatively small numbers of features, including the Naval propulsion plants (NPP) data, the Concrete compressive data and the Vehicle silhouette data. Table~\ref{table:compExhaust} summarizes the results. It can be seen that exhaustive search is only feasible for dataset with a very small number of features. 
Despite a much shorter running time, {\it msB} achieves an AUP score very close to that obtained by exhaustive search. Note that 
due to the excessively long running time, we repeat the exhaustive search on the NPP and Vehicle dataset only for 10 runs. 
\begin{table}
\begin{center}
\begin{tabular}{r|rrr}
\hline
   \bf{Dataset}     		&\#\bf{features}               		& \bf{Exhaustive search}     	& \bf{msB}   	\\[1pt]
\hline \\[-10pt]
Concrete       		&8		&28 (0.6975)					&14 (0.6904)					\\
NPP  		&16		&105657 (0.9781)   	     			&412 (0.9723)        			\\
Vehicle			&18		& 36143 (0.7672)				&27 (0.7559)				\\[1pt]
\hline
\end{tabular}
\caption{A comparison of exhaustive search and {\it msB} on running time (in seconds) and the resulting AUP score
(indicated in the parenthesis).} 
\label{table:compExhaust}
\end{center}
\end{table}
\section{Conclusions}
\label{section:conclusions}
This work tackles the challenging problem of measures or variables selection under a budget (or under different budget levels). 
We proposed an efficient strategy in navigating the solution space by following multiple model sequences with each having the 
potential of leading to an `optimal' solution. Instead 
of delivering a single model as output, we produce a model schedule which would allow a user to pick the model with the best 
predictive accuracy under a given budget, or to get the best tradeoff between model cost and predictive accuracy. Experiments 
on several benchmark or real datasets show that our approach compares favorably to competing methods. It would be interesting 
to explore different cost profiles, however, we feel that it is a big topic worth an extended study, and due to page limits, we decide 
to leave that to future work.
\\
\\
Given the high cost in the collection, storage, processing, and maintenance of large scale data, methods that incorporate variable 
costs will be highly desirable and widely applicable. We expect our approach could be generalized to many settings beyond classification. 
Also, our idea of multi-path search, along with the learning of model power from data and multicore computing
has the potential of becoming a general strategy for efficiently finding an approximate solution to a large class of discrete 
optimization problems.


\end{document}